\documentclass[article]{elsarticle}
\usepackage{graphicx}  
\usepackage{dcolumn}   
\usepackage{bm}        
\usepackage{verbatim}   
\journal{Physics letter A}
\newcommand{\F}[2]{\frac{#1}{#2}}
\newcommand{\D}{\partial}
\newcommand{\FD}[3]{\left(\frac{\D #1}{\D #2}\right)_{#3}}
\newcommand{\SD}[3]{\left(\frac{\D^2 #1}{\D #2^2}\right)_{#3}}
\newcommand{\BE}{\begin{equation}}
\newcommand{\EE}{\end{equation}}
\newcommand{\eq}[1]{~(\ref{#1})}

\begin{document}
\begin{frontmatter}
\title{General and exact pressure evolution equation}
\author{Adrien Toutant}
\ead{adrien.toutant@univ-perp.fr}
\address{
   University of Perpignan\\
   PROMES-CNRS (UPR 8521)\\
   Tecnosud, Rambla de la Thermodynamique\\
   66100 Perpignan, France
   }
\date{\today}
\begin{abstract}
A crucial issue in fluid dynamics is related to the knowledge of the fluid pressure. A new general pressure equation is derived from compressible Navier-Stokes equation. This new pressure equation is valid for all real dense fluids for which the pressure tensor is isotropic. It is argued that this new pressure equation allows unifying compressible, low-Mach and incompressible approaches. Moreover, this equation should be able to replace the Poisson equation in isothermal incompressible fluids. For computational fluid dynamics, it can be seen as an alternative to Lattice Boltzmann methods and as the physical justification of artificial compressibility. 
 \end{abstract}
\end{frontmatter}

\section{Introduction}
Incompressible Navier-Stokes equations (INS) describes a fluid characterized by infinite sound speed. It is valid in the case of fluid flows in isothermal configuration and at low Mach numbers (Mach number, $Ma=U^\star/c^\star$ is the ratio of the characteristic flow speed $U^\star$ and the speed of sound $c^\star$ defined at some reference temperature $T^\star$ and density $\rho^\star$). INS equations correspond to a mixture of hyperbolic and elliptic partial differential equations. They can be written
\BE
\D_tu_i + u_j\D_j u_i+\D_i P= \F{1}{Re}\D_j \D_j u_i~~~~~\D_i u_i=0\label{INS}
\EE
where $\mathbf{u}$ is the fluid velocity, $P$ is the pressure and $Re$ the Reynolds number, which represents the ratio between inertial and viscous forces~\cite{landau:1987}.  The pressure in~(\ref{INS}) is not an independent thermodynamic variable. It can be seen as a Lagrangian multiplier of the incompressibility constraint. It is determined by the Laplace or Poisson equation:
\BE
\D_i \D_i P=-(\D_j u_i)(\D_i u_j)\label{poisson}
\EE
In very anisothermal flow, the low Mach number hypothesis conducts to a similar system~\cite{paolucci:1982,toutant:13}. Considering that only density $\rho$ depends on temperature, the low Mach number equations can be written
\begin{eqnarray}
\D_tu_i + u_j\D_j u_i+\F{1}{\rho}\D_i P&=& \F{1}{\rho Re}\left(\D_j \D_j u_i+\F{1}{3}\D_i S\right)\nonumber\\
\D_i u_i&=&S\label{lowMach}
\end{eqnarray}
where $\rho$ depends on temperature and $S$ is a source term linked to conductive heat transfer ($S$ depends on temperature). Again, the pressure is determined by a Poisson equation. It can be given by:
\begin{eqnarray}
\D_i \left(\F{1}{\rho}\D_i P\right)&=&\F{1}{Re}\F{4}{3\rho}\D_j \D_j S+\nonumber\\
&&\F{1}{Re}\D_j\left(\F{1}{\rho}\right) \left(\D_i\D_i u_j+\F{1}{3}\D_i S\right)\nonumber\\
&&-(\D_j u_i)(\D_i u_j)-u_j\D_j S-\D_tS\label{poissonLM}
\end{eqnarray}
The physical meaning of~(\ref{poisson}) and~(\ref{poissonLM}) is that in a system with infinitely fast sound propagation, any pressure disturbance induced by the flow is instantaneously propagated into the whole domain. This elliptic problem is a crucial issue for fluid dynamics. Indeed, the INS equations are difficult to study analytically and numerically. This difficulty has motivated the search of alternative numerical approaches to determine pressure without solving the Poisson equation. Three different ways have been found. The first is the so-called artificial compressibility method where a pressure evolution equation is postulated~\cite{chorin:1967}. The second way is the Lattice Boltzmann method (LBM) which uses a velocity-space truncation of the Boltzmann equation from the kinetic theory of gases~\cite{benzi:1992}. The third way consists in adopting an inverse kinetic theory which permits the identification of the (Navier-Stokes) dynamical system and of the corresponding evolution operator which advances in time the kinetic distribution function and the related fluid fields~\cite{tessarotto:2008}. The pressure evolution equation obtained by this method is non-asymptotic. The full validity of INS equations is preserved.

In this paper, we determine a general and exact pressure evolution equation for all real dense fluids for which the pressure tensor is isotropic. Unlike the work of Tessarotto \textit{et al.}~\cite{tessarotto:2008}, the obtained pressure equation is a physical one and not a mathematically rigorous theory for INS equations. The obtained general and exact pressure evolution equation gives the physical bases of artificial compressibility method and it allows the study of very anisothermal flow contrary to LBM. The goal is similar to the reduced compressible Navier-Stokes equations (RCNS) derived by Ansumali \textit{et al.}~\cite{ansumali:2005} and the proposed pressure equation is very similar to the grand potential equation derived by Karlin \textit{et al.}~\cite{karlin:2006}. However we will argue that the use of pressure instead of the grand potential simplifies the equation of compressible hydrodynamics. Moreover, because the proposed pressure equation is valid for all real dense fluids, it builds bridge between compressible, low-Mach and incompressible approaches. 

In section 2, we will determine the general and exact pressure evolution equation (without any additional assumptions). This equation generalizes the one used by Zang \textit{et al.} in the particular case of an ideal gas~\cite{zang:1992}. In section 3, we will simplify this equation in the low Mach number limit. Finally, in section 4, we will reduce the equation for low Mach number and isothermal flow. 

\section{General pressure evolution equation}
The total energy $E$ conservation is given by
\BE
\D_t(\rho E)+\D_j ((\rho E+p)u_j)=\D_j(\sigma_{ij}u_i)-\D_j q_j
\EE
with $\sigma_{ij}$ the shear-stress tensor for a Newtonian fluid
\BE
\sigma_{ij}=2\mu S_{ij}-\F{2}{3}\mu\delta_{ij}S_{kk} ~~~S_{ij}=\F{1}{2}(\D_i u_j+\D_j u_i)
\EE
and $q_i$ the conductive heat flux
\BE
q_i=-\kappa\D_i T
\EE
Introducing internal energy $U=E-\F{u_i u_i}{2}$ and enthalpy $H=U+\F{P}{\rho}$, one gets
\BE
\rho D_tH=D_tP-\D_i q_i+\Phi~~~~~~~~\Phi=\sigma_{ij}\D_i u_j
\EE
with $D_t$ the material derivative (total derivative~\cite{landau:1987}):
\BE
D_t\phi=\D_t\phi+u_i\D_i\phi
\EE
Using the relations of heat capacity at constant pressure $c_p$ and of the isobaric thermal expansion coefficient $\alpha$
\begin{eqnarray}
\left(\F{\D H}{\D T}\right)_P&=&c_p \nonumber\\
\left(\F{\D H}{\D P}\right)_T&=&\F{1}{\rho}(1-T\alpha)~~with~~\alpha=-\F{1}{\rho}\left(\F{\D \rho}{\D T}\right)_P
\end{eqnarray}
an alternative formulation can be derived for temperature 
\BE
\rho c_p D_tT=T\alpha D_tP-\D_i q_i+\Phi\label{temperature}
\EE
We propose to derive a new pressure equation from the temperature formulation~(\ref{temperature}). We introduce the isothermal compressibility coefficient $\chi_T=\F{1}{\rho}\FD{ \rho}{P}{T}$ and we consider temperature as a function of density and pressure $D_tT=\FD{T}{\rho}{P}D_t\rho+\FD{T}{P}{\rho}D_tP$. Using mass conservation and~(\ref{temperature}), the recomputation poses no difficulties and we here write the result:
\BE
(\rho c_p \F{\chi_T}{\alpha}-\alpha T)D_tP=-\D_i q_i-\F{\rho c_p}{\alpha}\D_i u_i+\Phi\label{pressure1}
\EE
In order to simplify this expression, one introduces isochoric heat capacity $c_v=\FD{U}{T}{\rho}$, heat capacity ratio $\gamma=\F{c_p}{c_v}$ and the Mayer relation
\BE
\alpha^2 T=\rho c_v\chi_T(\gamma-1)\label{mayer}
\EE
One obtains
\BE
D_tP+\F{\gamma}{\chi_T}\D_i u_i=\F{\alpha}{\rho c_v\chi_T}\left(\Phi-\D_i q_i\right)
\EE
Sound velocity $c$ and the isentropic compressibility coefficient $\chi_S$ are given by
 \begin{eqnarray}
c^2&=&\FD{P}{\rho}{S} \nonumber\\
\chi_S&=&\F{1}{\rho}\FD{\rho}{P}{S}
\end{eqnarray}
Using the Reech relation $\gamma=\F{c_p}{c_v}=\F{\chi_T}{\chi_S}$, one obtains the general and exact pressure evolution equation
\BE
D_tP+\rho c^2 \D_i u_i=\F{\alpha}{\rho c_v\chi_T}\left(\Phi-\D_i q_i\right) \label{pressure}
\EE
In the particular case of an ideal gas $\alpha=\F{1}{T}$, $\chi_T=\F{1}{P}$ and $c^2=\gamma r T$ with $r$ the specific gas constant, this equation is equivalent to the one used by Zang \textit{et al.}~\cite{zang:1992,garnier:2009}. It is worth noting that equation~(\ref{pressure}) can be used for any real dense fluids (gas or liquid) without restriction on Mach number or temperature gradient. It gives the physical bases of artificial compressibility methods that postulate the pressure equation. The pressure equation~(\ref{pressure}) can be seen as an energy equation: to complete the system, one has to consider in addition mass conservation, momentum conservation and an equation of state. 

\section{Pressure equation for low Mach number flow}
At this step, the pressure evolution equation~(\ref{pressure}) depends on the total derivative.  At low Mach number, the first simplification consists in assuming that viscous dissipation is negligible. We now show that, at low Mach number, advection can be neglected. One defines the following nondimensionalized quantities:
\begin{eqnarray}
\rho^X=\F{\rho}{\rho^\star} &~~~~~~~&  u_i^X=\F{u_i}{c^\star}\nonumber\\
P^X=\F{\gamma P}{\rho^\star (c^\star)^2} & ~~~~~~~ & t^X_P=\F{1}{Ma^2}\F{tU^\star}{x^\star}\label{adim}
\end{eqnarray}
It is worth noting that in the classical low Mach number assumption, nondimensionalized time is defined by $t^X_U=\F{tU^\star}{x^\star}$. The factor $\F{1}{Ma^2}$ is justified by the fact that pressure time evolution is much faster than velocity time evolution (subscripts $_U$ and $_P$ indicate that the nondimensionalized time corresponds to velocity or pressure respectively). One defines moreover, the Reynolds number $Re$, the Prandtl number $Pr$ and the Peclet number $Pe$:
\BE
Re=\F{\rho U^\star x^\star}{\mu}~~~Pr=\F{\nu}{a_T}~~~Pe=PrRe
\EE 
One uses the asymptotic expansion of pressure, temperature and velocity
\begin{eqnarray}
P^X&=&P_0+Ma^2P_1 \\
T^X&=&T_0+Ma^2T_1 \\
u_i^X&=&Ma(u_{i0}+Ma^2u_{i1})
\end{eqnarray}
where $P^X$ is the nondimensionalized pressure, $P_0$ the zero-order pressure that is constant in space and $P_1$ the second-order pressure (see~\cite{paolucci:1982,majda:85}). It follows that the orders of magnitude of the different terms  of equation~(\ref{pressure}) are given by:
\begin{eqnarray}
\D_tP&=&\F{\rho^\star}{x^\star\gamma} (c^\star)^3\F{1}{Ma}\D_t(P_0+Ma^2P_1)\\
u_i\D_iP&=&\F{\rho^\star}{x^\star\gamma} (c^\star)^3Ma^3u_{i0}\D_iP_1\\
\rho c^2\D_iu_i&=&\F{\rho^\star}{x^\star} (c^\star)^3Ma\D_iu_{i0} \\
\F{-\alpha}{\rho c_v\chi_T}\D_i q_i&=&\F{\rho^\star}{x^\star} (c^\star)^3\F{Ma}{RePr}\D_i\D_iT_0
\end{eqnarray}
Consequently, the leading order gives:
\BE
\D_tP_0=0
\EE
The zero-order pressure $P_0$ is the reference pressure that is constant in time. It is worth noting that in the classical low Mach number assumption the zero-order pressure $P_0$ is not constant in time. This difference between the classical low Mach number assumption and the new low Mach number assumption proposed in this paper is due to the fact that asymptotic expansion depends of the chosen nondimensionalized quantities. More precisely, the zero-order pressure of this new low Mach number assumption is constant in time because of the choice of characteristic time for pressure variation. This choice is justified by the fact that pressure has very fast time variation compared to velocity for small Mach number. The second-order pressure $P_1$ corresponds to the thermodynamic pressure and the pressure of the momentum equation. The time evolution of the second-order pressure is given by:
\BE
\D_t(P_1)+\gamma\D_iu_{i0}=\F{\gamma}{RePr}\D_i\D_iT_0
\EE
The realized asymptotic expansion conducts to neglect advection for small Mach number. This hypothesis is further confirmed using the results of large eddy simulations and direct numerical simulations of biperiodic turbulent channel flow at different Reynolds numbers and different temperature ratios. The simulations are realized with the ideal gas assumption. The temperature ratio is defined as the ratio of wall imposed temperature at the hot wall to the wall imposed temperature at the cold wall (temperatures are in Kelvin). The studied temperature ratios are 2 and 5. The studied turbulent Reynolds numbers (based on the friction velocity) are 180 and 395.  More details about the simulations can be find in ~\cite{serra:12a,serra:12b,serra:12c,toutant:13,sanchez:14,aulery:2016}. In all these cases, at a given distance from the wall, the statistic average of the term  $\rho c^2 \D_i u_i=\gamma P\D_i u_i$ is a thousand times bigger than the statistic average of $u_i\D_iP$. Written with dimensionalized variables, the pressure evolution equation becomes for low Mach number (viscous dissipation is assumed to be negligible):
\BE
\partial_tP+\rho c^2 \D_i u_i=\F{-\alpha}{\rho c_v\chi_T}\D_i q_i\label{pressureLM}
\EE
In the particular case of an ideal gas, one finds the equation used for $P_0$ in the low Mach number approximation~\cite{paolucci:1982,majda:85}. We show here that equation~(\ref{pressureLM}) can be used for the pressure of momentum equation $P_1$ and for all fluids (gas and liquid). It is a local nonadvected equation for the scalar thermodynamic field. Combined with mass conservation, momentum conservation and an equation of state, it constitutes a system very similar to Lattice Boltzmann method (LBM): low Mach number equations avoiding the nonlocality of pressure (there is no Poisson equation and so no need for elliptic solver). Compared to LBM, the system is valid without the ideal gas assumption and the small temperature gradient restriction. Indeed, it is well known that LBM has stability problems  in very anisothermal flows. Compared to fully compressible Navier-Stokes, the energy equation is expressed in a pressure form without the ideal gas assumption and without the advection term.  

\section{Pressure equation for isothermal low Mach number flow}
At this stage, the pressure evolution equation~(\ref{pressureLM}) depends on temperature. We now consider the isothermal limit of this equation. Considering temperature as a function of pressure and density, one gets:
\begin{eqnarray}
\D_i\D_iT&=&\SD{T}{P}{\rho}(\D_iP)(\D_iP)+\SD{T}{\rho}{P}(\D_i\rho)(\D_i\rho)+\nonumber\\
&&\FD{T}{P}{\rho}\D_i\D_iP+\FD{T}{\rho}{P}\D_i\D_i\rho
\end{eqnarray}
For sufficiently large time-space scales (see the discussion in~\cite{ansumali:2005}), one can neglect density variation ($\rho$, $\chi_T$ and $\alpha$ are supposed contant and evaluated at equilibrium) and temperature variation (around a globally uniform equilibrium temperature) becomes a function of pressure:
\BE
\D_i\D_iT\approx\FD{T}{P}{\rho}\D_i\D_iP=\F{\chi_T}{\alpha}\D_i\D_iP
\EE 
It is worth noting that the pressure in this equation corresponds to $P_1$ the second-order pressure. This pressure corresponds to the pressure of the momentum equation ($P_0$ the zero-order pressure is constant in space). In the isothermal limit, thermal conductivity is assumed to be constant: $-\D_iq_i=\kappa\D_i\D_iT$. The isothermal pressure evolution (IPE) equation is finally given by:
\BE
\partial_tP+\rho c^2 \D_i u_i=\F{\kappa}{\rho c_v}\D_i\D_iP\label{pressureiso}
\EE
Note that all ``material parameters'' appearing in\eq{pressureiso} ($\kappa$, $\rho$ and $c_v$) are evaluated at a constant equilibrium temperature. A very interesting point of this equation is the presence of a diffusion term. By analogy with temperature diffusivity $a_T=\F{\kappa}{\rho c_p}$, one can define the pressure diffusivity $a_P=\F{\kappa}{\rho c_v}$. The ratio of pressure/temperature diffusivities corresponds to the heat capacity ratio $\gamma=\F{a_p}{a_T}$. This diffusion term is a crucial difference with artificial compressibility method~\cite{chorin:1967}. This term allows to stabilize the simulation. It guarantees the numerical applicability of the method. Indeed, Ohwada and Asinari~\cite{ohwada:2010} proposed to introduce a dissipation term in order to improve the quality of numerical solution obtained with artificial compressibility method. Moreover, the kinetically reduced local Navier-Stokes (KRLNS) equations~\cite{karlin:2006} contains exactly the same diffusion term applied to grand potential instead of pressure. In fact, the KRLNS grand potential equation is exactly the same than the proposed IPE equation\eq{pressureiso} if one replaces grand potential ${\cal G}$ by pressure $P$. This observation leads to two conclusions. In one hand, it is expected that numerical simulations using IPE equation associated with momentum equation capture the correct transient behavior of complex flows as KRLNS does~\cite{hashimoto:2015}. On the other hand, because KRLNS grand potential equation and IPE equation are the same, kinetic energy  $k=P-{\cal G}$ can be neglected in KRLNS. Indeed, the substraction of the KRNLS grand potential equation from the proposed IPE equation\eq{pressureiso} gives
\BE
\partial_tk=\F{\kappa}{\rho c_v}\D_i\D_ik
\EE
In the case of a Prandtl number different from heat capacity ratio ($Pr\neq\gamma$), this last equality will not be accurate unless kinetic energy and its space/time variations are very small. 

Finally, it is interesting to write the IPE equation\eq{pressureiso} in a nondimensional form. Omitting the exponent $^X$ for the sake of simplicity, the nondimensionalized form of\eq{pressureiso} is 
\BE
\D_tP+\F{\gamma}{Ma^2}\D_i u_i= \F{\gamma}{RePr}\D_i\D_iP
\EE
It shows that ``pressure diffusion'' cannot be neglected for pressure time evolution because its order of magnitude can be similar to the one of pressure time derivative due to the Peclet number. Again, it is a fundamental difference with artificial compressibility method. It allows to stabilize the simulation. Furtheremore, considering a Reynolds number around one and a Prandtl number similar to the Mach number squared (very small Prandtl number), pressure diffusion has the same order of magnitude of the divergence term. Consequently, it is anticipated that isothermal low Mach number flows may not be incompressible. To the best of our knowledge, experimental evidence of compressible ($\D_iu_i\neq 0$) isothermal low Mach number flows does not exist. It would be very interesting to investigate. Considering numeric interest of IPE, because the Prandtl number has no effect on isothermal incompressible flow, it is expected that it can be used in the IPE equation\eq{pressureiso} as a numerical parameter to stabilize the simulation. Obviously, the Peclet number has to be much bigger to the Mach number if one wants to approximate the incompressible limit. 

\section{Conclusion}
This work establishes the general and exact pressure equation evolution. The pressure equation has been obtained
\begin{itemize}
\item for real dense fluids for which the pressure tensor is isotropic,
\item in the low Mach number limit and  
\item in the isothermal limit. 
\end{itemize} 
For real dense fluids for which the pressure tensor is isotropic, the obtained equation generalizes the ideal gas pressure equation. In the low Mach number limit, we propose a new asymptotic analysis that conducts to neglect advection. The proposed low Mach number assumption conducts to the same equation than the classical one. However, the new equation corresponds to the pressure of momentum equation. It means that the coupling between the energy equation and the momentum equation is increased. Consequently, the new assumption is less restrictive than the classical one. It is expected that it better corresponds to the physics of turbulent anisothermal flows where turbulent time scales can be of the same order of magnitude as pressure time scales. In the isothermal limit, the pressure equation gives the physical bases of ACM. The obtained equation is different from ACM. Indeed, a new diffusion term is added. This new term is coherent with the fact that both viscosity and thermal conductivity are involved in acoustic wave attenuation. 

In a fundamental point of view, this equation provides a thermodynamic theory of incompressible hydrodynamics. More precisely, it gives an asymptotic\footnote{A pressure evolution equation directly (not asymptotically) consistent with the isochoricity condition should probably follow from a first-principle microscopic/kinetic statistical description.} thermodynamic derivation of incompressibility without the isentropic flow assumption. Such a thermodynamic derivation goes far beyond academic interest. Indeed, the general pressure evolution equation paves the way for new technique or methodology for computational fluid dynamics. In future works, we  will use this equation as an alternative numerical approach to determine pressure without solving the Poisson equation as ACM or LBM.

\bibliographystyle{elsarticle-num}

\bibliography{biblio}

\end{document}